\documentstyle[aps,prl,epsf,floats]{revtex}
\begin{document}
\twocolumn[\hsize\textwidth\columnwidth\hsize\csname@twocolumnfalse%
\endcsname
\title{Quantum spin glass transition in the two-dimensional electron gas}
\author{Subir Sachdev}
\address{Department of Physics, Yale University, P.O. Box 208120, New
Haven, CT 06520-8120}
\date{September 17, 2001}

\maketitle

\begin{abstract}
We discuss the possibility of spin glass order in the vicinity of
the unexpected metallic state of the two-dimensional electron gas
in zero applied magnetic field. An average ferromagnetic moment
may also be present, and the spin glass order then resides in the
plane orthogonal to the ferromagnetic moment. We argue that a
quantum transition involving the destruction of the spin glass
order in an applied in-plane magnetic field offers a natural
explanation of some features of recent magnetoconductance
measurements. We present a quantum field theory
for such a transition and compute its mean field properties.~\\
~\\
\end{abstract}
]

The pioneering experiments of Kravchenko, Pudalov, Sarachik, and
others~\cite{rmp} have demonstrated that the low density electron
gas in two dimensions (the 2DEG), in the absence of an applied
magnetic field, exhibits some fascinating strong correlation
physics. The initial indications of this new regime came from an
unexpected decrease in the resistivity as the temperature was
lowered on the electrons in a silicon inversion layer.
Experimentally it is of course impossible to definitively rule out
that there will be an eventual upturn of this resistivity at some
exponentially small temperature, and the that the true ground
state is always an insulator. Nevertheless, the
single-electron-transistor measurements of Ilani {\em et
al.}~\cite{amir} show that there is a remarkable qualitative
change in the fluctuations of the local compressibility at the
same densities, and there must be a corresponding transition in
the quantum ground state. A theoretical description of this
correlation-induced transformation remains an important
theoretical challenge.

Useful guidance is obtained from numerical studies of the
disorder-free 2DEG at the same densities. These indicate that the
system is not too far from a Wigner crystallization transition,
and that non-trivial spin correlations are present~\cite{bernu}. A
study of an effective spin Hamiltonian~\cite{sudip,misguich} for
the Wigner crystal ground state shows instabilities to
ferromagnetism~\cite{cc} and to a spin singlet ground state which
is an attractive candidate for fractionalization.

Turning to the disordered 2DEG, we can reasonably expect that
there will be regions of the sample where the system is locally
ferromagnetic\cite{aleiner}. However, in other regions the ring
exchange terms may prefer a spin singlet state, and these should
couple the different ferromagnetic regions so that they align in
different directions. On these grounds, we propose here that the
ground state has spin glass order in the vicinity of the
transition in the conductivity, possibly with a concomitant
average ferromagnetic moment; if the ferromagnetic moment is
non-zero, the spin glass order resides in a plane (in spin space)
orthogonal to the direction of the moment. We expect that the spin
glass character is stronger on the metallic side, while the
ferromagnetic character of the ground state becomes stronger as
the density of electrons is reduced, and this is partially
responsible for the decrease in the conductivity. Other arguments
for the proximity of the spin glass order, which were based
directly on the theory of the disordered metal, were presented
in~\cite{ssroyal}.

Our proposal is also motivated by recent experimental studies of
the magnetization of the 2DEG~\cite{sergey,krav,pud}. In
particular, the measurements of Vitkalov~{\em et
al.}~\cite{sergey} in an in-plane (``parallel'') magnetic field,
$H$, indicate that there is a well-defined critical field
$H_{\sigma}$, dependent on the density of the 2DEG, at which the
magnetoconductance displays a kink-like feature, and which they
identify with a quantum phase transition. For a clean 2DEG it is
natural to interpret $H_{\sigma}$ as a saturation field, above
which all electrons have their spin polarized in the direction of
the applied field. In a disordered 2DEG there is no reason for all
regions of the sample to respond in the same manner, and there
will always be local configurations in which the electrons prefer
to form low spin clusters at these relatively weak fields.
Naively, one would then conclude that the magnetization of a
disordered 2DEG should evolve smoothly as a function of the
applied field, without non-analytic structure at any critical
field $H_{\sigma}$. A simple way out of this dilemma is to search
for some other order parameter associated with the preserved
symmetry of spin rotations about the direction of the applied
field. Spin glass order in the plane perpendicular to the magnetic
field is our proposed candidate: this order is present for $H <
H_{\sigma}$, but vanishes for $H > H_{\sigma}$ when the
polarization of many spins along the direction of the applied
field reduces their ``canting'' enough to disrupt temporal memory
of their orientation in the orthogonal plane.

(The discussion of this present paper will restrict attention to
the case where $H_{\sigma} > 0$. Experimentally, it is known that
$H_{\sigma}$ decreases as the density of electrons decreases, and
a possible reason for this is the increase in the spontaneous
ferromagnetic moment which acts as an effective magnetic field. It
is possible that there is a density at which $H_{\sigma}$ reaches
zero: this corresponds to a transition at $H=0$, driven by tuning
the density, between a ferromagnetic state with spin glass order
in the transverse plane and an ordinary ferromagnet. The theory
for this density-tuned transition will be similar to the
field-tuned transition described below, but will not be presented
here: it follows by combining the results below with those
of~\cite{rsy,ss,pd}.)

The presence of a critical point at $H=H_{\sigma}$ also leads to a
natural explanation of another puzzling feature of the data. For
the density at which $H_{\sigma}$ was small, it was found that the
characteristic $H$-width of the magnetoconductance feature scaled
roughly as the absolute temperature, $\sim k_B T$ (we have
absorbed a factor of $g \mu_B$ in the definition of $H$, where
$\mu_B$ is the Bohr magneton and $g$ is the $g$-factor of
electrons in the silicon conduction band.) A trivial system whose
thermodynamics depends only on $H/T$ are isolated free spins.
However given the large exchange interaction energies between the
electrons, it appears extremely unlikely that there is a
sufficient density of isolated free moments to lead to a
significant change in the magnetoconductance. Indeed there is a
`catch-22' here: if the moments are really isolated enough to
behave as free spins, their coupling to the itinerant electrons is
weak and they have a negligible effect on the conductance. The
presence of a spin-glass quantum critical point at $H=H_{\sigma}$
offers an alternative route to obtaining a characteristic field of
order $T$. Indeed it can be argued on rather general grounds that
if the quantum critical point obeys the hyperscaling property, and
if $H$ couples to a conserved total spin, then the characteristic
field scale will be of order $k_B T$~\cite{conserve}.

It is appropriate to now mention very interesting recent
observations of glassy behavior, persisting in the metallic phase,
by Bogdanovich and Popovi\'{c}~\cite{drag}, which appear to be
consistent with our proposal. These authors made the connection to
glassy behavior in models of charge transport of spinless
electrons~\cite{dob}, while here we will focus only on the spin
degrees of freedom.

We will now propose a quantum field theory for the transition at
$H=H_{\sigma}$. We will follow general approach to quantum spin
glasses developed in~\cite{rsy} and reviewed in Chapter 16
of~\cite{book}. In the presence of an applied magnetic field,
there is only a U(1) symmetry of spin rotations about the
direction of the applied field, and we are interested in
singularity in the dependence of the conserved ``charge'' of this
symmetry (the magnetization) on the applied field. A general
theory for such transitions was presented in~\cite{sss} and in
Chapter 11 of~\cite{book}. Here we will extend this theory to
random systems with spin glass order by combining it with the
methods of~\cite{rsy}.

Let us assume that the field $H$ is applied in the $z$ direction,
and let $S_{\alpha} (r,\tau)$ with $\alpha=x,y$ be the component
of the electron spin density in the orthogonal $x$,$y$ plane at
the spatial point $r$ and at imaginary time $\tau$. We use the
standard replica method to treat the quenched disorder:
consequently, we introduce replica indices $a=1,\ldots,n$, and the
spin density $S_{\alpha a} (r, \tau)$. The quantum theory for spin
glass order is expressed in terms of the order parameter
functional $Q_{\alpha\beta}^{ab} (r, \tau_1, \tau_2)$ which is
\begin{equation}
Q_{\alpha\beta}^{ab} (r, \tau_1 , \tau_2 ) \sim S_{\alpha a} (r,
\tau_1) S_{\beta b} (r, \tau_2 ). \label{qsg7}
\end{equation}
By applying the methods developed in~\cite{rsy,book} to spin glass
order in the $x$,$y$ plane in the presence of an applied magnetic
field we obtain the following low order terms in the effective
action for $Q_{\alpha\beta}^{ab} (r, \tau_1, \tau_2)$:
\begin{eqnarray}
&& {\cal S}_{sg} =  \int d^d r \left\{ \frac{1}{\kappa} \int d\tau
 \sum_a \Biggl[ \frac{1}{2}\left(-i
\frac{\partial}{\partial\tau_1}+i \frac{\partial}{\partial \tau_2}
\right) \epsilon_{\alpha\beta} \right.
\nonumber \\
&&~~~~~~~~+ \left. \left(H-H_{\sigma}^0
\right)\delta_{\alpha\beta} \Biggr] Q^{aa}_{\alpha\beta} (x ,
\tau_1 , \tau_2 )
\right|_{\tau_1=\tau_2=\tau}  \nonumber \\
&&~~~~~~~~+ \frac{1}{2} \int  d \tau_1 d \tau_2 \sum_{ab} \left[
\nabla Q_{\alpha\beta}^{ab} (r, \tau_1, \tau_2 )
\right]^2  \nonumber \\
&&- \frac{\kappa}{3} \int  d \tau_1 d \tau_2 d \tau_3 \sum_{abc}
Q^{ab}_{\alpha\beta} (r, \tau_1 , \tau_2 ) Q^{bc}_{\beta\rho} (r,
\tau_2 , \tau_3 ) Q^{ca}_{\rho\alpha}
(r, \tau_3 , \tau_1 ) \nonumber \\
&&~~~~~~~~~  + \frac{1}{2} \int  d \tau \sum_a \left[ u~
Q^{aa}_{\alpha\beta} ( r, \tau , \tau) Q^{aa}_{\alpha\beta} ( r,
\tau , \tau) \right.\nonumber \\
&&~~~~~~~~~~~~~\left. +v~ Q^{aa}_{\alpha\alpha} ( r, \tau , \tau)
Q^{aa}_{\beta\beta} ( r, \tau , \tau) \right]\Biggr\},
\label{landau}
\end{eqnarray}
where $d=2$ is the spatial dimensionality, summation of the spin
indices $\alpha,\beta,\rho$ over $x,y$ is implied,
$\epsilon_{\alpha\beta}$ is the antisymmetric tensor,
$H_{\sigma}^0$ is the mean-field value of the critical field
(which will be renormalized by fluctuations), $\kappa, u, v$ are
coupling constants, and we have omitted an additional quadratic
term which has no influence on the mean-field theory, but does
play an important role in the violation of hyperscaling in the
perturbative fluctuations~\cite{rsy}. In a metal with an
appreciable contribution of low energy spin excitations associated
with the particle-hole continuum, there will also be additional
dissipative term in the action, as discussed in~\cite{sro}: in the
present situtation with strong local correlations we believe this
is unlikely to be the case, and so have omitted such a term above.

An important property of (\ref{landau}) is that the field $H$
couples to a conserved U(1) charge: consequently changes in $H$
can be absorbed by a time-dependent gauge transformation which is
equivalent to a transformation into a ``rotating reference
frame''~\cite{conserve,book}. More specifically, if we generalize
the action to a time-dependent field $H(\tau)$, then it is
invariant under the infinitesimal transformation
\begin{eqnarray}
Q_{\alpha\beta}^{aa} (r, \tau_1, \tau_2) &\rightarrow&
Q_{\alpha\beta}^{aa} (r, \tau_1, \tau_2) - \epsilon_{\alpha\gamma}
\phi (\tau_1) Q_{\gamma\beta} (r, \tau_1, \tau_2) \nonumber \\
&~&~~~~~~~~~~~~~~- \epsilon_{\beta\gamma} \phi (\tau_2)
Q_{\alpha\gamma} (r, \tau_1,
\tau_2) \nonumber \\
H(\tau) &\rightarrow& H(\tau) - i \partial_{\tau}\phi (\tau),
\label{gauge}
\end{eqnarray}
where $\phi (\tau)$ is infinitesimal. If the critical point at
$H=H_{\sigma}$ satisfies strong hyperscaling properties, then
(\ref{gauge}) implies that the free energy density ${\cal F}_{sg}$
obeys
\begin{equation}
{\cal F}_{sg} = {\cal F}_0 + T^{\mu} \Phi \left(
\frac{H-H_{\sigma}}{T^{\varphi}} \right) \label{fsg}
\end{equation}
where the exponents $\mu=1+d/z$, $\varphi=1$, $z$ is the dynamic
critical exponent, and $\Phi$ is a scaling function. Taking the
$H$ derivative of (\ref{fsg}) we obtain the magnetization
\begin{equation}
M = M_0 - T^{\mu-\varphi} \Phi^{\prime} \left(
\frac{H-H_{\sigma}}{T^{\varphi}} \right) \label{fsg1}
\end{equation}
where $M_0$ is the background ferromagnetic magnetization present
for $H \gg H_{\sigma}$. The onset of spin glass order causes a
decrease in this magnetization as $H$ is lowered. A related
scaling form should also hold for the magnetoconductance.

We illustrate the above behavior of the magnetization by a simple
mean field analysis of ${\cal S}_{sg}$ in (\ref{landau}).
Unfortunately these mean field results do not satisfy hyperscaling
properties appropriate to any value of $d$, and the reasons for
this are similar to those discussed at length in~\cite{rsy} for
other quantum spin glasses. The actual situation for realistic
spin glasses remains an open problem, as an analysis of
fluctuations about the mean field solutions leads to a runaway
flow to strong coupling. If our proposal is indeed the correct
explanation for the experimental data~\cite{sergey}, then the
hyperscaling prediction $\varphi=1$ must be valid for $d=2$.

At the mean-field saddle-point, we may make the following ansatz
for $Q_{\alpha\beta}^{ab} (r, \tau_1, \tau_2)$ based upon the
requirements of translational invariance in space and time (we
henceforth set $k_B = 1$):
\begin{equation}
Q_{\alpha\beta}^{ab} (r, \tau_1, \tau_2) = T \sum_{\omega_n}
\widetilde{D}_{\alpha\beta}^{ab} ( \omega_n ) e^{-i \omega_n
(\tau_1 - \tau_2)}, \label{g1}
\end{equation}
where $\omega_n$ is a Matsubara frequency, and
\begin{equation}
\widetilde{D}_{\alpha\beta}^{ab} (\omega_n) = \delta_{ab}
D_{\alpha\beta} (\omega_n) + \delta_{\alpha\beta}
\frac{\delta_{\omega_n,0}}{T} q_{EA}. \label{g2}
\end{equation}
The first, replica-diagonal term in (\ref{g2}) is the local
dynamic spin susceptiblity in the $x$,$y$ plane, while $q_{EA}$ is
the Edwards-Anderson spin glass order parameter. For simplicity we
have assumed a replica-symmetric form for the spin glass
order---replica symmetry is not broken for the terms included in
(\ref{landau}), but is broken when higher order terms are
included: this is as discussed in~\cite{rsy}. Notice also that the
second term in (\ref{g2}) also includes a contribution of $q_{EA}$
along the replica diagonal---this is unlike the usual procedure in
the theory of classical spin glasses; the diagonal contribution
here accounts for the long-time limit of the local spin
correlation function. Further analysis is simplified by rewriting
the $D_{\alpha\beta}$ in ``circularly-polarized'' components:
\begin{eqnarray}
D_{xx} = D_{yy} &=& \frac{1}{2} \left( D_{+-} + D_{-+} \right)
\nonumber \\
D_{xy} = -D_{yx} &=& \frac{i}{2} \left( D_{+-} - D_{-+} \right).
\label{g3}
\end{eqnarray}

Inserting (\ref{g1}-\ref{g3}) into (\ref{landau}) we obtain the
following expression for the free energy density after taking the
limit $n \rightarrow 0$:
\begin{eqnarray}
{\cal F}_{sg} &=& {\cal F}_0 + \frac{T}{\kappa} \sum_{\omega_n}
\Bigl\{ (-i \omega_n + H - H_{\sigma}) D_{+-} (\omega_n)
\nonumber\\ &+& (i \omega_n + H - H_{\sigma}) D_{-+} (\omega_n) +2
q_{EA} (H - H_{\sigma}) \Bigr\} \nonumber \\ &-& \frac{\kappa
T}{3} \sum_{\omega_n} \Bigl\{ D^3_{+-} (\omega_n) + D^3_{-+}
(\omega_n) \Bigr\} \nonumber \\ &-& \kappa q_{EA} \Bigl\{D^2_{+-}
(0) + D^2_{-+} (0)\Bigr\} \nonumber \\
&+& u  \left[ q_{EA} + T \sum_{\omega_n} D_{+-} (\omega_n) \right]
\left[ q_{EA} + T \sum_{\omega_n} D_{-+} (\omega_n) \right]
\nonumber \\ &+& \frac{v}{2} \left[ 2 q_{EA} + T \sum_{\omega_n}
\Bigl\{ D_{+-} (\omega_n) + D_{-+} (\omega_n) \Bigr\} \right]^2;
\label{g4}
\end{eqnarray}
we have replace $H_{\sigma}^0$ by $H_{\sigma}$ anticipating the
mean-field position of the critical point. It is now
straightforward to determine the saddle-point of ${\cal F}_{sg}$
with respect to variations in $D_{+-} ( \omega_n)$, $D_{-+}
(\omega_n)$, and $q_{EA}$. We find two classes of solutions
describing the paramagnetic and spin glasses phases respectively,
and we will discuss their properties in turn.

In the paramagnetic phase we have $q_{EA} = 0$, and
\begin{eqnarray}
D_{+-} (\omega_n) &=& - \frac{1}{\kappa} \sqrt{- i \omega_n + \Delta} \nonumber \\
D_{-+} (\omega_n) &=& - \frac{1}{\kappa} \sqrt{ i \omega_n +
\Delta} \label{g5}
\end{eqnarray}
where the energy $\Delta$ is determined by the solution of
\begin{equation}
\Delta = H-H_{\sigma} - (u+2v) T\sum_{\omega_n} \sqrt{- i \omega_n
+ \Delta}. \label{g6}
\end{equation}
The condition $\Delta \geq 0$ delineates the boundary of the
paramagnetic phase. At $T=0$, the Matsubara summation in
(\ref{g6}) becomes a frequency integral which evaluates to zero
(the expressions in (\ref{g5},\ref{g6}) only aim to capture the
singular low frequency behavior, and it is assumed that the high
frequency form is such that contours of frequency integration can
be freely closed at complex infinity). So $\Delta=H-H_{\sigma}$ at
$T=0$, which demonstrates that the paramagnetic phase is stable
only for $H>H_{\sigma}$. For $T>0$, (\ref{g6}) can be analyzed
using methods that have been discussed in some detail in Chapter
15 of~\cite{book}; for $H-H_{\sigma}$ small, its solution can be
written as
\begin{equation}
\Delta + (u+2v) T \sqrt{\Delta} = H-H_{\sigma} + (u+2v) T^{3/2}
\Xi \left( \frac{H-H_{\sigma}}{T} \right), \label{g7}
\end{equation}
with the function $\Xi$ is given by
\begin{equation}
\Xi (y) = \frac{1}{\pi} \int_0^{\infty} \sqrt{\Omega} d \Omega
{\cal P} \left(\frac{1}{e^{\Omega+y} - 1}\right) +
\theta(y)\sqrt{y}, \label{g8}
\end{equation}
where ${\cal P}$ denotes a principle part and $\theta(y)$ is the
unit step function. The expression (\ref{g8}) was derived for
$y>0$ ($H>H_{\sigma}$) but has been written in a manner which
defines it for real $y$. Despite appearances, the function $\Xi
(y)$ is actually analytic at $y=0$, and indeed it is analytic at
all real values of $y$; for small $y$, $\Xi (y) = \zeta (3/2)/(2
\sqrt{\pi}) +  0.411958y + \ldots$. For $y<0$, the results
(\ref{g7},\ref{g8}) apply in the paramagnetic portion of the phase
diagram present at $T>0$, $H<H_{\sigma}$ (see Fig.~\ref{fig1}
below). At the critical field, $H=H_{\sigma}$, (\ref{g7}) predicts
that $\Delta = (u+2v) \Xi (0) T^{3/2}$; application of
hyperscaling to this quantum critical region would have implied
$\Delta \sim T \times \mbox{a function of $(H-H_{\sigma})/T$}$,
and so it is evident that hyperscaling is not obeyed by the
mean-field theory. Imposing the condition $\Delta=0$ in (\ref{g7})
determines the mean-field boundary of the spin glass phase as
$H=H_{\sigma} - (u+2v) \Xi (0) T^{3/2}$ at small $T$: this leads
to the phase diagram in Fig~\ref{fig1}--the structure of the
crossovers is very similar to those discussed earlier for other
spin glasses, and the reader is referred to Chapter 15
of~\cite{book} for a review.
\begin{figure}[htbp]
\epsfxsize=8cm \centerline{\epsfbox{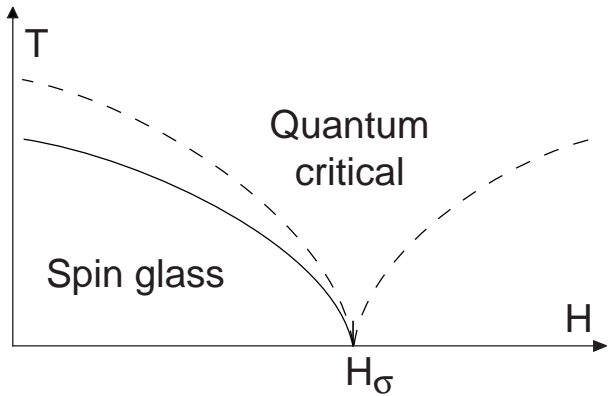}} \caption{Phase
diagram of the quantum spin glass described by ${\cal S}_{sg}$ in
(\protect\ref{landau}). The full line is a phase transition
marking the mean-field boundary of the spin glass phase--in $d=2$
spin glass order may only be present at $T=0$. The dashed lines
are crossovers into the quantum-critical region, which survives
fluctuations in $d=2$. With hyperscaling, the characteristic
energy in the quantum-critical region $\sim T$; the mean-field
theory has this energy $\sim T^{3/2}$.} \label{fig1}
\end{figure}
\begin{figure}[htbp]
\epsfxsize=8cm \centerline{\epsfbox{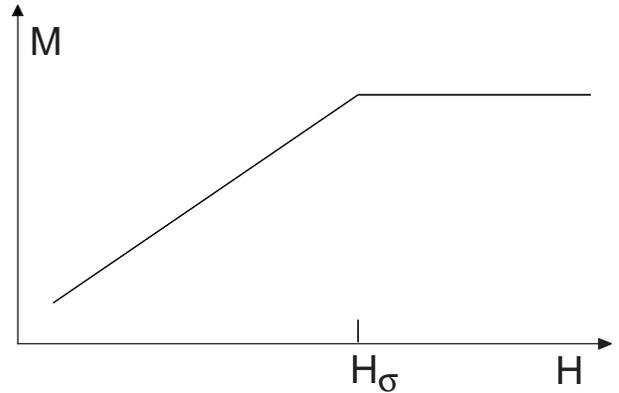}} \caption{Mean field
prediction for the magnetization, $M$, as a function of $H$ at
$T=0$ in (\ref{g9a}) and (\ref{g11a}). Upon including
fluctuations, the singularity at $H=H_{\sigma}$ will survive but
the magnetization will be $H$ dependent for $H>H_{\sigma}$. At
$T>0$ the singularity at $H=H_{\sigma}$ will be rounded out on a
scale $T$ if hyperscaling is obeyed by the critical point.}
\label{fig2}
\end{figure}
Note that the paramagnetic phase extends to $H<H_{\sigma}$ for
$T>0$, and that the expression (\ref{g7}) remains valid in this
regime where $y<0$. The free energy in the paramagnetic phase is
also easily obtained from (\ref{g4}-\ref{g6}), and we obtain
\begin{equation}
{\cal F}_{sg} = {\cal F}_0 - \frac{(\Delta - H +
H_{\sigma})^2}{\kappa^2 (u + 2v)} - \frac{4 T^{5/2}}{3 \pi
\kappa^2} \int_0^{\infty} \frac{ \Omega^{3/2} d
\Omega}{e^{(\Omega+\Delta)/T} - 1} \label{g9}
\end{equation}
The $T$ and $H$ dependence of the magnetization is determined by
taking a $H$ derivative of (\ref{g9}):  we find
\begin{equation}
M=M_0~~~\mbox{as $T \rightarrow 0$ for $H> H_{\sigma}$},
\label{g9a}
\end{equation}
up to exponentially small terms, indicating that the magnetization
is effectively saturated in the paramagnetic phase in this
mean-field theory, as indicated in Fig~\ref{fig2}. Fluctuations
will induce a more appreciable variation in the magnetization even
at $T=0$, as strict saturation is not possible in a disordered
system.

Finally, we describe the saddle point of (\ref{g4}) in the
spin-glass phase. Here we find the simple solution
\begin{eqnarray}
D_{+-} (\omega_n) &=& - \frac{1}{\kappa} \sqrt{- i \omega_n } \nonumber \\
D_{-+} (\omega_n) &=& - \frac{1}{\kappa} \sqrt{ i \omega_n}
\nonumber \\
q_{EA} &=& \frac{H_{\sigma}-H}{\kappa (u+2v)} + \frac{T}{\kappa}
\sum_{\omega_n} \sqrt{-i \omega_n} \nonumber \\
&=& \frac{H_{\sigma}-H}{\kappa (u+2v)} - \frac{\Xi (0)
T^{3/2}}{\kappa}
\label{g10}
\end{eqnarray}
The last expression identifies the same boundary of the spin glass
phase (where $q_{EA}=0$) as that determined above. The free energy
can also be computed as before, and we obtain
\begin{equation}
{\cal F}_{sg} = {\cal F}_0 - \frac{(H_{\sigma} - H)^2}{\kappa^2 (u
+ 2v)} - \frac{4 T^{5/2} \Gamma(5/2) \zeta(5/2)}{3 \pi \kappa^2},
\label{g11}
\end{equation}
and the magnetization follows as its $H$ derivative. Now we find
\begin{equation}
M=M_0 - \frac{2 (H_{\sigma} - H)}{ \kappa^2 (u+2v)}~~~\mbox{as $T
\rightarrow 0$ for $H< H_{\sigma}$}. \label{g11a}
\end{equation}
Comparing with (\ref{g9a}) we see that there is a kink in the
magnetization at the quantum critical point $H=H_{\sigma}$, as
shown in Fig~\ref{fig2}. This singularity survives fluctuation
corrections, even though the mean-field saturation of the
magnetization in (\ref{g9a}) does not.

This papers has outlined a scenario by which the 2DEG can exhibit
a quantum phase transition at a critical in-plane applied magnetic
field $H=H_{\sigma}$: the transition is induced by the destruction
of spin-glass order in the plane orthogonal to the applied field.
We have argued that the instabilities of the spin exchange
Hamiltonian for the ordered Wigner crystal to ferromagnetism and
to spin-singlet states lend support to the possibility of
spin-glass order in the disordered 2DEG. It would be interesting
the compare the $H$, $T$, and density dependent data to scaling
forms like (\ref{fsg},\ref{fsg1}). Theoretical analysis of the
transport properties of the field-tuned transition at
$H=H_{\sigma}$, and also of the density-tuned transition for the
case $H_{\sigma}=0$, will also be of interest.

\acknowledgements I thank Sergey Vitkalov, Myriam Sarachik, and
Sergey Kravchenko for useful discussions. This research was
supported by US NSF Grant DMR 0098226.

\end{document}